\documentclass[rnote]{aa}

\begin{document}
\title{FH Leonis, the first dwarf nova member of a multiple star system?}
\subtitle{}
\author{Nikolaus Vogt\inst{1},\inst{2}}
\offprints{N. Vogt, \email{nvogt@ucn.cl}}
\institute{Instituto de Astronomia, Universidad Catolica del Norte, Antofagasta, Chile
 \and     Departamento de Fisica y Astronomia, Universidad de Valparaiso, Valparaiso, Chile}
\date{Received  / Accepted }
\abstract
  {}
  {Sudden brightenings of  HD 96273 or BD+07 2411B were observed with the HIPPARCOS satellite in 1992,
    which still require a convincing explanation.}
  {A new analysis of all known data of these two stars is given, including additional
   information on the Balmer line equivalent widths.}
  {The brightenings can be explained as SU UMa type dwarf nova outbursts, superimposed on the combined
  light of two normal F and G type main sequence stars. Since the hypothetical dwarf nova turns
   out to be located at the same distance as HD 96273 and BD+07 2411B, we possibly see here the
   first case of a cataclysmic variable as a member of a multiple star system.  Questions on history
   and evolution, as well as possible ways to confirm this interpretation, are briefly outlined.}
  {}
\maketitle

\section{Introduction}
\label{sec:introduction}

A recent paper in this journal, by Dall et al. (\cite{dall}, hereafter DSSI) was titled
``Outbursts on normal stars -- FH Leo misclassified as a novalike variable''.
The authors report three possible brightenings of the combined flux of the wide
binary HD 96273 and BD+07 2411B observed with the HIPPARCOS satellite in 1992,
January 3, January 15 and June 19 with amplitudes between $0^{m}.1$ and $0^{m}.35$ above
the normal brightness level of this double star. They obtained high resolution
spectra of both stellar components, determined their effective temperatures,
surface gravities, microturbulence parameters, rotation and radial velocities as
well as abundances of several chemical elements. Since both stars behave
--according to all these parameters-- like normal main sequence dwarfs, DSSI concluded that
the identification as a cataclysmic variable must be wrong. They discussed
several possible causes of the observed brightenings, in particular transient background
or foreground objects, magnetic interaction with an unseen companion, a planetary
accretion event and a microlensing event. Finally, they point out that none of
these explanations seems to be convincing.

In this Research Note I would like to show that most of the observed properties
of FH Leo can have their natural explanation if we assume that it is a dwarf nova
at the same distance as  HD 96273 and BD+07 2411B.

\section{FH Leo as a dwarf nova companion of HD 96273 or BD+07 2411B}
\label{sec:fhleo}

The HIPPARCOS distance of this binary is 117 pc. This implies
an absolute magnitude M$_{V}$ = $+3^{m}.6$ of HD 96273 and M$_{V}$ = $+4^{m}.8$
of BD+07 2411B. The effective temperatures determined
by DSSI correspond to spectral types F6V  (HD96273)  and G3V
(BD+07 2411B). These absolute magnitude values and also the
observed colours are in accordance with a classification of
a normal main sequence star, the interstellar reddening can
be neglected. The combined absolute magnitude in the blue spectral
range is M$_{B}$ = $3^{m}.76$. Since on 1992, January 3 (JD 2448624) a brightening
of $0^{m}.27$ on  average  was observed, the absolute magnitude
of the total light during this event was M$_{B}$ = $+3^{m}.49$.
Therefore, the absolute magnitude of the brightening
source alone is M$_{bs}$ = $+5^{m}.13$, which is nearly identical of that of a dwarf nova
at outburst maximum: equation (3.4) of Warner (\cite{warner}) predicts
a value between $+5^{m}.2$ and $+5^{m}.3$ for a dwarf nova with an orbital
period between 1.5 and 2 hours. On 1992, January 15 (JD 244
8636) the absolute magnitude of the brightening source was M$_{bs}$ = $+6^{m}.1$,
about $1^{m}$  fainter than 12 days before. A natural
explanation of this behaviour is that of an SU UMa type dwarf
nova which begun its superoutburst in 1992, January 3 and declined
slowly towards the end of its plateau phase in January 15. This
way, even the differences of about $0^{m}.1$  between the brighter
and the fainter magnitude measures in January 3 could be explained
by the superhump activity; the brighter points may refer to
the superhump peaks, their time difference is about 2 hours,
compatible with the superhump period of a typical SU UMa
type dwarf nova. The total intrinsic superhump amplitude would
be of the order of $0^{m}.4$, just as expected during the early
stage of a superoutburst.

One important question, however, remains,  the absorption spectra
of HD 96273 and BD+07 2411B look rather normal. In particular,
DSSI did not find any traces of hydrogen emission which would be typical
for dwarf novae in quiescence. For instance, Thorstensen et
al. (\cite{thorstensen}) give H$\gamma$ emission equivalent widths of 40\AA\ for V844
Her and 20\AA\ for DI UMa. In H$\beta$ even values up to 100\AA\ are possible
(Warner, \cite{warner}, Fig. 2.35). Indeed, the hydrogen emission may
be present in FH Leo, but unobservable due to the large difference
in brightness between HD 96273/BD+07 2411B and the hypothetical
dwarf nova in quiescence. If  FH Leo is an SU UMa type dwarf
nova with a long outburst cycle, similar to WX Cet (Sterken
et al., \cite{sterken}) we expect an amplitude $\ge6^{m}$ , revealing an absolute
magnitude $\ge11^{m}.1$ of the dwarf nova in quiescence.
Because of the large difference in the continuum flux, even
in the blue spectral range, an emission equivalent width of
H$\gamma$ of 40\AA\ of the quiescent dwarf nova would be reduced to a
contribution of about 0.045\AA\ in the equivalent widths of HD
96273 or BD+07 2411B while the above maximal EW value of H$\beta$
would contribute 0.11\AA\. DSSI did not publish their
measured equivalent widths of these stars, but the authors kindly sent me an
original spectrum of each star. From this I have determined the
equivalent widths of H$\alpha$, H$\beta$, H$\gamma$ and H$\delta$. Their
values are $3.7 \pm 0.2$, $5.9 \pm 0.3$, $7.4 \pm 0.8$ and $5.1 \pm 0.3$ resp. for HD 96273 and
$3.0 \pm 0.2$, $3.6 \pm 0.3$, $5.5 \pm 1.0$ and $3.2 \pm 0.2$ resp. for BD+07 2411B (the given errors refer
mainly to uncertainties in the continuum determination.
An emission contribution of the dwarf
nova in quiescence would reduce such equivalent widths by a
very small amount ($\le$2\%) which certainly is impossible to recognize
even in high resolution spectroscopy. Therefore, it is not surprising
that DSSI did not find any anomaly in the spectra of HD 96273
and BD+07 2411B.

The brightening event of 1992, June 19 (JD 2448793), about 169
days after the ``superoutburst'' detection, could perhaps be explained
as a short outburst of the dwarf nova. However, there are some
other measurements without a significant brightening, only a
few hours before and afterwards. Therefore, this event could
also refer to a short-lived ``flare'' which also was observed
in other dwarf novae (Bateson \cite{bateson}). In any case, further observations
will be necessary to confirm the existence of short eruptions
or flares in FH Leo.

\section{Conclusions}
\label{sec:conclusions}

The most plausible explanation of the outburst behavior of FH
Leo would be that of an SU UMa type dwarf nova, superimposed
on the wide binary HD 96273 and BD+07 2411B, and placed at
the same distance as this binary. If this is true we either
have a very rare and improbable chance coincidence, or simply
a gravitationally bound multiple star system with a dwarf nova
orbiting HD 96273 or BD+07 2411B. To my knowledge, this would
be the first case of the detection of a cataclysmic variable
as component of a multiple star system. Rather interesting questions
would arise concerning the possible evolutionary history of
such a system: What is the age of it? What was its origin and how did it
evolve? Since the HD 96273/BD+07 2411B system is a rather wide
binary (projected physical separation 936 AU) there is plenty
of space for hierarchical sub-structures of smaller dimensions,
such as a giant-type progenitor of a cataclysmic variable which
has to pass the subsequent common envelope phase.

This interpretation, however, needs to be confirmed. This can
be done in different ways: First of all, we will monitor FH
Leo in the next season closely and search for additional outbursts,
which also should reveal superhumps, the main characteristic
of any SU UMa type dwarf nova. This will be done in our Cerro
Armazones Observatory (Universidad Catolica del Norte, Antofagasta)
and possibly at other institutions. On the other hand, we will
search for outbursts of FH Leo in the patrol plate archive of
the Sonneberg Observatory, which is, following Harvard, the
world-wide second largest in size and plate number (Braeuer and
Fuhrmann \cite{brauer}). It covers several decades beginning about 1930,
and was used to detect more than 10000 new variable stars.
The typical threshold of detection with blink comparator or
similar visual inspection methods, as applied traditionally
at the Sonneberg Observatory, is about $0^{m}.3$, just the total
amplitude of FH Leo. Therefore, it is not surprising that FH
Leo remained undetected until recently. New scanning and reduction
methods as applied by Vogt et al. (\cite{vogt}), however, allow us to reduce
this threshold to less than $0^{m}.1$, thus enabling the detection
of historical outbursts of FH Leo. A search for them is in preparation.

\section{Acknowledgements}

I would like to thank Dr. Thomas H. Dall for kindly providing me the
original spectra, and to Dr. Eduardo Unda-Sanzana for his assistance
in the data treatment. I also acknowledge the support of FONDECYT (grant 1061199).

\end{document}